\documentstyle[multicol,prb,aps,graphics,epsf]{revtex}

\begin{document}

\def\bb{\begin{equation}}
\def\ee{\end{equation}}

\title{Low energy valence photoemission in the Anderson impurity model for
Ce compounds}
\author{J.D. Lee}
\address{Max-Planck Institut f\"{u}r Festk\"{o}rperforschung, \\
Heisenbergstrasse 1, D-70569 Stuttgart, Germany}
\date{\today}
\maketitle

\begin{abstract}
The valence level photoemission spectra in the Anderson impurity model 
for Ce compounds at zero temperature are studied as a function of
the photon energy $\omega$.
Most of former studies on Ce compounds are based on the sudden
approximation, which is valid in high energy region.
For the photoemission in the adiabatic limit of low energy region,
one should consider the dipole matrix elements and the hole-induced 
photoelectron scattering potential.
We can manage it by combining the time-evolution
formalism and the $1/N_f$ scheme in a large $f$-level degeneracy $N_f$.
This gives the exact results as $N_f\rightarrow\infty$.
In view of experiments on the valence photoemission, two contributions of
$4f$- and band emissions are mixed. We study the separate $4f$ and band
contributions (from Ce $5d$) 
and total emission including the interference between two 
on an equal footing with varying the photon energy.
In the $4f$-emission case, we also explore the effects of hole-induced 
scattering potential of the photoelectron with respect to $\omega$.
Its effects are found very similar to the core level photoemission in
shake down case with a localized charge transfer excitation.
Additionally, we examine the adiabatic-sudden transition
in valence level photoemission for the present localized system 
through the simplified two-level model.
\end{abstract}

\begin{multicols}{2}

\section{Introduction}

The Anderson impurity model (AIM) was originally proposed to discuss
the property of magnetic impurities in nonmagnetic metals\cite{Anderson}.
After that, AIM has been widely applied to the analysis of 
spectroscopic data for $f$ and $d$ electron systems, i.e. 
rare earth compounds\cite{Kotani} or transition metal compounds\cite{Sawat}, 
where electron states are treated to be an impurity
and they are hybridized with the valence or conduction electron states.
Also, AIM has been often used to describe the 
Ce mixed-valence compounds, where one considers the $f$ level on one atom
and its interaction with the conduction bands.
In investigations of Ce and its compounds, the basic question
concerns the nature of a $4f$ electron and other electronic states and
how they mix with the $4f$ state. Much of the interests are therefore 
imposed on the properties of the $4f$ states, i.e. the occupation,
position, width, coupling to the metallic band, intra-atomic Coulomb
interaction and so on. 
There were numerous studies of thermodynamic and transport 
properties for them, which has been followed by the electron-spectroscopy 
studies\cite{Lawrence81}.

The photoemission spectroscopy (PES) is a very useful tool 
for studying the electronic structure of matters and 
could have provided a lot of insights also for Ce studies.
But it's worth noting PES cannot always give 
a simple answer about the underlying electronic structure
because the photoelectron may perturb the system left behind. 
Actual description of theoretical PES is quite complicated and therefore
the sudden approximation is frequently used, where the photoelectron
is assumed decoupled from the remaining solid\cite{Hedinbook}.
The sudden approximation becomes exact when the kinetic energy
of the emitted electron gets large infinitely.

Gunnarsson and Sch\"{o}nhammer\cite{Gunnar83} have extensively studied the 
electron spectroscopies for Ce compounds, i.e. core level photoemission,
x-ray absorption, and bremsstrahlung isochromat spectroscopy
as well as the valence photoemission. For the information of the position 
and width of $f$-level in the compounds, the valence photoemission has often
been used. Through their studies of valence photoemission in
$f$-emission channel, they reproduced two-peak structure
in Ce compound using AIM consistent with the experiments. 
In the earlier evolution stage for Ce materials, it was found that 
the valence PES shows just a single $f$-related structure 
2-3 eV below the Fermi level\cite{Allen,Croft}.
Later PES experiments have demonstrated 
the $4f$ spectrum has the additional structure interestingly 
near the Fermi level\cite{Marten,Peter,Wiel,Suga}.
Subsequently, it was shown that the particular structure is due to
the Kondo resonance singlet characterized by the small energy $T_K$. 
In the actual experiments on valence photoemission, 
two contributions of $4f$-emission and band emission 
(from Ce $5d$ or other bands) are mixed.
The identification of $4f$-emission from the experiments
is a highly nontrivial work. 
Wieliczka et al.\cite{Wiel} have done the comparison 
of spectra taken at the different photon energies showing the
resonance of $4f$-emission\cite{Lawrence82} and reported the additional peak 
near the Fermi level. Another possibility is to assume
the behaviors of $4f$- and band emissions with respect to the
photon energies, especially in $20-80$ eV\cite{Patthey}.
Nevertheless, they could have said nothing about the interference between two.
These works can motivate to explore a more explicit analysis 
for the interference effects of two emission channels
with the photon energy varied.
Gunnarsson and Sch\"{o}nhammer\cite{Gunnar85} have also studied 
the band emission contribution as well as $4f$-emission and discussed 
the interference effects of two. However, all their works were
within the sudden approximation. 

We consider both the contributions of $4f$- and $5d$-emission
on an equal footing by introducing two dipole matrix elements,
$\Delta_f(E)$ and $\Delta_d(\epsilon,E)$, where $E$ is the 
kinetic energy of photoelectrons. In the sudden approximation, the dipole 
matrix elements are normally treated constant with $E$. 
But in the low energy PES, it can be crucial.
Combining the time-dependent formalism and $1/N_f$ idea, 
we can calculate the PES exactly up to ${\cal O}(\frac{1}{N_f})^0$
as the photon energy varies. $N_f=\infty$ can be a good
approximation to $N_f=14$ in Ce compounds.
Then we study the separate contributions of 
$4f$- and $5d$-emission and more interestingly the nontrivial 
interference effects between two. The relative sign or strength 
of $\Delta_f(E)$ and $\Delta_d(\epsilon,E)$ are important.
Difference in the energy scale of $\Delta_f(E)$ and $\Delta_d(\epsilon,E)$
makes the spectra from each channel separate with respect to $\omega$
in the low energy PES. It is also found that, because the interference
contribution has a peak near the Fermi level, the $4f$-derived 
peak near the Fermi level may be enhanced 
or suppressed in a total spectra.

It is recently reported that the adiabatic-sudden transition 
due to the photoelectron scattering potential will be governed 
by the characteristic of relevant excitations to which the emitted electron 
couples in the system. When the photoelectron couples to the
extended excitation like plasmon, the sudden transition occurs 
in very large kinetic energies ($\sim$ keV)\cite{Hedin98},
while to the localized excitations, the sudden approximation 
occurs much quicker\cite{Lee}. 
In our study, the extrinsic effects beyond the sudden approximation
will be considered only through the hole-induced scattering potential.
Effects like surface or several damping mechanisms 
will not be taken into account.
The system of Ce compounds has also included the localized excitations 
represented by $f^0$, $f^1$, $f^2$ created from the hole potential.
We can consider the scattering potential in the valence PES due to $f$-hole
in the impurity model within the formalism. The scattering effects 
will not be important for band emission channel.
The crossover from adiabatic to
sudden limit can be also reexamined in this localized system, for which 
we simplify the model to have only two relevant levels. The same criterion
for the transition is found as in the previous work\cite{Lee},
the energy scale of $\tilde{E}=1/(2\tilde{R}^2)$, 
where $\tilde{R}$ is a scattering potential range.

We organize the paper as follows. Our model and the formalism for calculation
are given in Sec. II. The simple sudden approximation results 
for separate and both channels are described in Sec. III.
In Sec. IV, we present the model for necessary 
dipole matrix elements and calculate the spectra for the separate 
contributions for $4f$- and $5d$-emission and for both with respect to
the photon energies. We also discuss
the interference contribution between two.
In Sec. V, for $4f$-emission, we include the photoelectron scattering
potential within the same formalism. In Sec. VI, we try to reexamine 
the adiabatic-sudden crossover in the system 
by way of the simplified two-level model. In Sce. VII, we 
give the discussion and conclusion.

\section{Model and Formalism}

As mentioned in the introduction, we consider the AIM Hamiltonian in the
energy basis used in Gunnarsson and Sch\"{o}nhammer's 
discussion for Ce compounds\cite{Gunnar83,Gunnar85},
\begin{eqnarray}\label{II1}
{\cal H}&=&{\cal H}_0+\Delta,
\nonumber\\
{\cal H}_0&=&\sum_{\nu}\int E\psi_{E\nu}^{\dagger}\psi_{E\nu}dE
            +\sum_{\nu}\int\epsilon\psi_{\epsilon\nu}^{\dagger}
             \psi_{\epsilon\nu}d\epsilon
            +\epsilon_f\sum_{\nu} n_{\nu}
\nonumber\\
          &+&\sum_{\nu}\int V(\epsilon)(\psi_{\nu}^{\dagger}
                            \psi_{\epsilon\nu}
                            +\psi_{\epsilon\nu}^{\dagger}\psi_{\nu})d\epsilon
          +\frac{U}{2}\sum_{\nu\neq{\nu}^{\prime}}n_{\nu}n_{{\nu}^{\prime}},
\end{eqnarray}
where $\psi_{E\nu}^{\dagger}(\psi_{E\nu})$ is a photoelectron operator,
$\epsilon$ denotes the $5d$ conduction states, $\epsilon_f$ describes
the impurity $4f$ level, and $V(\epsilon)$ is a hybridization matrix element
between the conduction states and localized $f$ level. 
$|V(\epsilon)|^2$ can be modelized to have a semielliptical
form symmetric with respect to $\epsilon_F=0$,
$\pi|V(\epsilon)|^2=2V^2(B^2-\epsilon^2)^{1/2}/B^2$, 
where $2B$ is the bandwidth. 
$\Delta$ in the Hamiltonian is the dipole term describing the
photon-matter interaction.
The one particle basis used in Eq.(\ref{II1}) is introduced
by assuming\cite{Bringer}
$$
\sum_{\bf k}V_{{\bf k}m}^{\ast}V_{{\bf k}m^{\prime}}
=|V(\epsilon)|^2\delta_{mm^{\prime}},
$$ 
$$
\psi_{\epsilon\nu}^{\dagger}=V(\epsilon)^{-1}\sum_{\bf k}V_{{\bf k}m}^{\ast}
                  \delta(\epsilon-\epsilon_{\bf k})
                  \psi_{{\bf k}\sigma}^{\dagger},
$$
and so $\nu$ is the orbital and spin magnetic quantum number and
from $\nu=1$ to $\nu=N_f$ if we assume the magnetic degeneracy $N_f$
of $f$-level. In Ce, $N_f$ is normally taken as 14.
To apply $1/N_f$ idea, we need one subsidiary condition that
$N_f^{1/2}V(\epsilon)$ should be independent of $N_f$.

In $\Delta$, we will generally have two interaction terms due to $4f$ level
and $5d$ conduction bands, so 
\begin{eqnarray}\label{II2}
\Delta&=&\sum_{\nu}\int dE\left[\Delta_{f}(E)\psi_{E\nu}^{\dagger}\psi_{\nu}
                     +\Delta_{f}^{\ast}(E)\psi_{\nu}^{\dagger}\psi_{E\nu}\right]
\\ \nonumber
      &+&\frac{1}{N_f^{1/2}}\sum_{\nu}\int d\epsilon dE
         \left[\Delta_{d}(\epsilon,E)\psi_{E\nu}^{\dagger}\psi_{\epsilon\nu}
         +\Delta_{d}^{\ast}(\epsilon,E)\psi_{\epsilon\nu}^{\dagger}\psi_{E\nu}
          \right].
\end{eqnarray}
By giving the explicit time-dependency $f(\tau)$ in $\Delta$ and redefining 
the dipole matrix elements, $\Delta_{f}(E)$ and $\Delta_{d}(\epsilon,E)$,
\begin{eqnarray}\label{II3}
\Delta_{f}(E)&\rightarrow&M_f\Delta_f(E)f(\tau),
\nonumber \\
\Delta_{d}(\epsilon,E)&\rightarrow&M_d\Delta_d(\epsilon,E)f(\tau),
\end{eqnarray}
\bb\label{II4}
f(\tau)=e^{-i\omega\tau}(e^{-\eta\tau}-1), \ \ \eta>0
\ee
we use a time-dependent formulation and solve the Schr\"{o}dinger equation for
the total Hamiltonian ${\cal H}$.

We first introduce a state $|0\rangle$,
\bb\label{II5}
|0\rangle=\prod_{\nu}^{N_f}\prod_{\epsilon<\epsilon_F}
          \psi_{\epsilon\nu}^{\dagger}|{\rm vac}\rangle,
\ee
where all the conduction electron states below Fermi energy
are occupied and the $f$ level is empty. For the simplicity, we keep
only the lowest order terms of $1/N_f$ before and after the photoemission,
which means the results will be exact as $N_f\rightarrow\infty$.
\bb\label{II6}
|\epsilon\rangle=\frac{1}{N_f^{1/2}}\sum_{\nu}\psi_{\nu}^{\dagger}
          \psi_{\epsilon\nu}|0\rangle,
\ee
\bb\label{II7}
|\epsilon,\epsilon^{\prime}\rangle
          =\frac{1}{N_f^{1/2}(N_f-1)^{1/2}}\sum_{\nu\neq\nu^{\prime}}
           \psi_{\nu}^{\dagger}\psi_{\nu^{\prime}}^{\dagger}
           \psi_{\epsilon^{\prime}\nu^{\prime}}\psi_{\epsilon\nu}|0\rangle,
\ee
\bb\label{II8}
|E,\epsilon\rangle=\frac{1}{N_f^{1/2}}\sum_{\nu}\psi_{E\nu}^{\dagger}
           \psi_{\epsilon\nu}|0\rangle,
\ee
\bb\label{II9}
|E,\epsilon,\epsilon^{\prime}\rangle=\frac{1}{N_f^{1/2}(N_f-1)^{1/2}}
          \sum_{\nu\neq\nu^{\prime}}
          \psi_{E\nu}^{\dagger}\psi_{\nu^{\prime}}^{\dagger}
          \psi_{\epsilon^{\prime}\nu^{\prime}}\psi_{\epsilon\nu}|0\rangle,
\ee
\begin{eqnarray}\label{II10}
|E,\epsilon,\epsilon^{\prime},\epsilon^{\prime\prime}\rangle
       &=&\frac{1}{N_f^{1/2}(N_f-1)^{1/2}(N_f-2)^{1/2}}
\nonumber \\
       &\times& \sum_{\nu\neq\nu^{\prime}\neq\nu^{\prime\prime}}
        \psi_{E\nu}^{\dagger}\psi_{\nu^{\prime}}^{\dagger}
        \psi_{\nu^{\prime\prime}}^{\dagger}
        \psi_{\epsilon^{\prime\prime}\nu^{\prime\prime}}
        \psi_{\epsilon^{\prime}\nu^{\prime}}
        \psi_{\epsilon\nu}|0\rangle.
\end{eqnarray}
Within the above basis set, after time $\tau$, the wave function
$|\Psi(\tau)\rangle$ of the system is given by
\begin{eqnarray}\label{II11}
|\Psi(\tau)\rangle&=&a(\tau)|0\rangle
          +\int b(\epsilon;\tau)|\epsilon\rangle d\epsilon
          +\int c(\epsilon,\epsilon^{\prime};\tau)
               |\epsilon,\epsilon^{\prime}\rangle d\epsilon d\epsilon^{\prime}
\nonumber \\
           &+&\int d(E,\epsilon;\tau)|E,\epsilon\rangle dEd\epsilon
\nonumber \\
           &+&\int e(E,\epsilon,\epsilon^{\prime};\tau)
            |E,\epsilon,\epsilon^{\prime}\rangle dEd\epsilon d\epsilon^{\prime}
\nonumber \\
           &+&\int f(E,\epsilon,\epsilon^{\prime},\epsilon^{\prime\prime};\tau)
             |E,\epsilon,\epsilon^{\prime},\epsilon^{\prime\prime}\rangle
             dEd\epsilon d\epsilon^{\prime}d\epsilon^{\prime\prime}.
\end{eqnarray}
The coefficients of $|\Psi(\tau)\rangle$ can be determined
by the time dependent Schr\"{o}dinger equation,
\bb\label{II12}
i\frac{\partial}{\partial\tau}|\Psi(\tau)\rangle
               ={\cal H}|\Psi(\tau)\rangle,
\ee
where the initial condition of the state should be corresponding to
the ground state before the photoemission, 
$|\Psi(\tau=0)\rangle=|\Psi_0\rangle$,
\bb\label{II13}
|\Psi(0)\rangle=a(0)|0\rangle
          +\int b(\epsilon;0)|\epsilon\rangle d\epsilon
          +\int c(\epsilon,\epsilon^{\prime};0)
           |\epsilon,\epsilon^{\prime}\rangle d\epsilon d\epsilon^{\prime}
\ee
and the equations for $a(0)$, $b(\epsilon;0)$, and 
$c(\epsilon,\epsilon^{\prime};0)$ are found in Ref.15.
The coefficients $M_f$ and $M_d$ represent 
the external field strength. In the present formalism, we solve 
the equation in the limit of $M_f\rightarrow 0$ and
$M_d\rightarrow 0$ and let the system evolve for a time of the order $1/\eta$ .
Then we can show the solution identical to the more conventional 
photoemission\cite{Lee,Gunnar80}. $\eta$ is a small positive number
and gives a life-time broadening in the spectra. In the actual calculation, 
$\eta$ is taken as 0.3 eV (0.01 au). The photoemission spectra
will now be proportional to
\begin{eqnarray}\label{II14}
I(E)&=&
\int|d(E,\epsilon;\tau)|^2 d\epsilon
+\int|e(E,\epsilon,\epsilon^{\prime};\tau)|^2 d\epsilon d\epsilon^{\prime}
\nonumber \\
& &+\int|f(E,\epsilon,\epsilon^{\prime},\epsilon^{\prime\prime};\tau)|^2
 d\epsilon d\epsilon^{\prime}\epsilon^{\prime\prime},
\end{eqnarray}
and we see, due to $M_f\rightarrow 0$ and $M_d\rightarrow 0$,
\bb\label{II15}
I(E)=\alpha(E)M_f^2+\beta(E)M_d^2+\gamma(E)M_fM_d,
\ee
where $\alpha(E)$, $\beta(E)$, and $\gamma(E)$ correspond to 
$4f$-, $5d$-emission, and interference between those, respectively. 

\section{Sudden approximation}

In the sudden approximation, we normally neglect $E$-dependency of
the dipole matrix elements, i.e. $\Delta_f(E)=M_f$ and
$\Delta_d(\epsilon,E)=M_d\Delta_d^{\prime}(\epsilon)$, 
where $E$ is a kinetic energy of photoelectron.
Gunnarsson and Sch\"{o}nhammer\cite{Gunnar85} assumed 
$\Delta_d^{\prime}(\epsilon)$ have the same shape as $V(\epsilon)$ 
for the conduction band emission, which we'll simply follow.

\begin{figure}
\vspace*{7.0cm}
\includegraphics{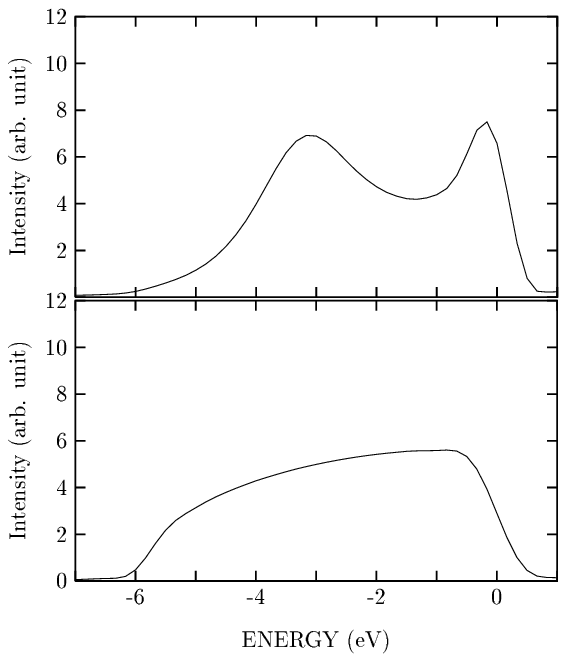}
\caption{In the upper panel, the $f$-derived valence PES is provided 
($M_d=0$) and in the lower panel, 
the conduction band emission is given ($M_f=0$)
for $U=5.0$ eV, $\epsilon_f=-2.5$ eV, $V=0.5$ eV, and $B=6$ eV. 
Both calculations are based on the sudden approximation.
Spectral curves are normalized to have the same area.
}
\label{fig:sud_separ}
\end{figure}

The AIM has often been studied in the limit of $U=\infty$, where 
it becomes so simple as to allow the analytic solutions. 
In our formalism, to neglect $|\epsilon,\epsilon^{\prime}\rangle$
and $|E,\epsilon,\epsilon^{\prime},\epsilon^{\prime\prime}\rangle$
corresponds to the limit.
However the assumption $U=\infty$ is not really justified 
because $U$ is just about $5\sim6$ eV\cite{Herbst}
and thus $f^0$ and $f^2$ configurations
are energetically comparable, i.e. $\epsilon_f$ is about 
$-2\sim -3$ eV\cite{Allen} and $2\epsilon_f+U\approx 0$.
In the calculations, we have always taken $U=5.0$ eV and 
$\epsilon_f=-2.5$ eV to be $2\epsilon_f+U=0$.
In Fig.\ref{fig:sud_separ}, we give the simple
valence PES results for $4f$- and $5d$-emission, respectively.

It is seen in Fig.\ref{fig:sud_separ} that we nicely reproduce
the well-known sudden $4f$-PES results\cite{Gunnar85} 
having the double-peak structure in the upper panel and also get 
$5d$-PES simulating the broad structureless conduction band.
The $4f$-PES is especially interesting because of its ample physics.
The peak well below the Fermi level corresponds to $4f$ ionization peak,
$4f^1\rightarrow 4f^0$, and the peak near the Fermi level 
(also called Kondo resonance peak) arise from 
a $4f$ hole screened by a $4f$ electron (making a $5d$ hole near 
the Fermi level), $4f^1\rightarrow 4f^1$.

In the separate calculations of emission channels, the absolute values 
or signs of $M_f$ and $M_d$ are surely irrelevant to the results.
There can be however some subtleties when we consider both 
emission channels. The PES curves drastically change with respect to
the relative sign or relative ratio of $M_f$ and $M_d$. 
For the relative strength, we parametrize the ratio of
$|M_d\Delta^{\prime}(0)/M_f|$ (note $|M_d\Delta^{\prime}(0)/M_f|>1$
does not always mean the band emission is dominant over the $f$-emission).
The relative sign is related to whether the interference will be
constructive or destructive.

\begin{figure}
\vspace*{9.5cm}
\includegraphics{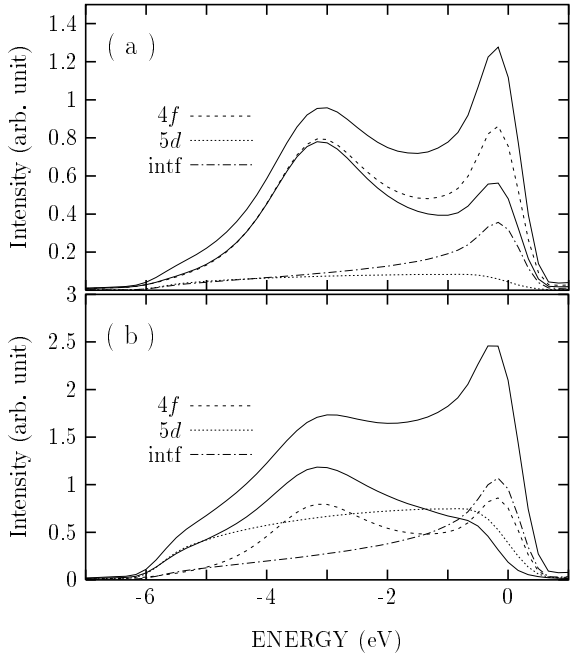}
\caption{The valence PES including both contributions of $4f$-
and $5d$-emission is shown. The solid line is the total spectra
corresponding to constructive or destructive interference, 
where interference effects (labeled "intf") is added or subtracted.
In (a), we use $|M_d\Delta^{\prime}(0)/M_f|=2$ and in (b),
$|M_d\Delta^{\prime}(0)/M_f|=6$. The used parameters are same in
Fig.1.
}
\label{fig:sud_total}
\end{figure}

As shown in Fig.\ref{fig:sud_total}, the relative ratio and signs 
of two dipole components are crucial in
valence PES. Interestingly, the interference contributions 
show a peak near the Fermi level, which may enhance the Kondo resonance 
peak from $4f$-emission in the constructive case 
or suppress in the destructive case.
In Fig.\ref{fig:sud_total}(b), 
even if we can see a clear ionization peak near $\epsilon_f$,
we see only the shoulder structure
not a peak near the Fermi level due to a strong destructive interference.
These destructive interferences may be one of the reasons that 
several groups in the former stage have failed to 
see a double-peak structure and reported only an ionization peak 
near $\epsilon_f$.
 
\section{Low energy valence photoemission: effects of dipole matrix}

In the last section, we have illustrated the sudden approximation results
valid in the high energy PES. To see how the PES behaves in the low energy
regime, first off we should account for the $E$-dependent dipole
matrix elements. Using the {\it Slater-type} orbital\cite{Slater}
for the corresponding atomic orbital, 
we calculate $E$-dependent dipole matrix elements.
The Slater orbital for $R_{nl}(r)$ is given by
\bb\label{IV1}
R_{nl}(r)=(2\zeta)^{n+1/2}[(2n)!]^{-1/2}r^{n-1}e^{-\zeta r},
\ee
where the orbital exponent is determined by a suitable rule.
But in $4f^1 5d^1 6s^2$ configuration of Ce, the Slater orbital for $5d$
gives actually the poor representation compared to a more
accurate LSD calculation\cite{Vosko} for the atomic wavefunction for Ce.
We adopt therefore the same functional form of Eq.(\ref{IV1}), but
determine the exponent $\zeta$ suitably by comparing with the 
accurate result, i.e. we take $\zeta_{4f}=5.0$ and $\zeta_{5d}=2.0$
(by a Slater-rule, $\zeta_{5d}$ will be 0.75). 
In principle, the atomic orbital and photoelectron basis function
having an explicit angular momentum channel $l$
should be obtained by solving the Schr\"{o}dinger equation
under the same Ce atomic potential.
But in our discussion the basis function is 
simply assumed to be a spherical Bessel function of $l$,
\bb\label{IV2}
\varphi_E^l(r)=\frac{\sqrt{2}}{\sqrt{\pi}}(2E)^{1/4}j_l(\sqrt{2E}r),
\ee
and its normalization follows
\bb\label{IV3}
\int r^2dr{\varphi_E^l}^{\ast}(r)
           \varphi_{E^{\prime}}^l(r)=\delta(E-E^{\prime}).
\ee
So the dipole matrix elements for $4f$-emission is given by
\bb\label{IV4}
\Delta_f(E)=M_f\int r^2dr R_{4f}(r)r\varphi_E^{l=4}(r),
\ee
and for the $5d$-conduction band emission, we assume $\Delta_d(\epsilon,E)$
has the simple separable form like
\bb\label{IV5}
\Delta_d(\epsilon,E)=M_d\Delta_d^{\prime}(\epsilon)\Delta_d(E).
\ee
$\Delta_d^{\prime}(\epsilon)$ is still assumed to have the shape 
of $V(\epsilon)$ as in the last section and $\Delta_d(E)$ 
can be expected to have much $5d$ atomic orbital character
if we think of the tight binding idea for the corresponding energy band.
Thus we assume the behavior of $\Delta_d(E)$ as
\bb\label{IV6}
\Delta_d(E)\propto\int r^2dr R_{5d}(r)r\varphi_E^{l=3}(r).
\ee
Here it should be noted that the possible $l$-channel of photoelectrons
are $l=2,4$ for $4f$-emission and $l=1,3$ for $5d$-emission due to
the angular momentum selection rule, but 
the major channel will be $l=4$ and $l=3$, respectively.

\begin{figure}
\vspace*{5.cm}
\includegraphics{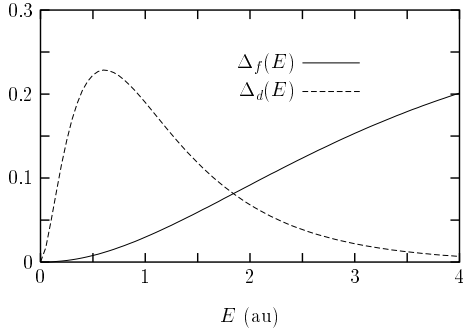}
\caption{The behavior of dipole elements - $\Delta_f(E)$, $\Delta_d(E)$ 
are provided with respect to the photoelectron kinetic energy $E$.
Note the different energy scale in two behaviors.
The absolute values are arbitrary.
}
\label{fig:dipol_E}
\end{figure}

\begin{figure}
\vspace*{9.5cm}
\includegraphics{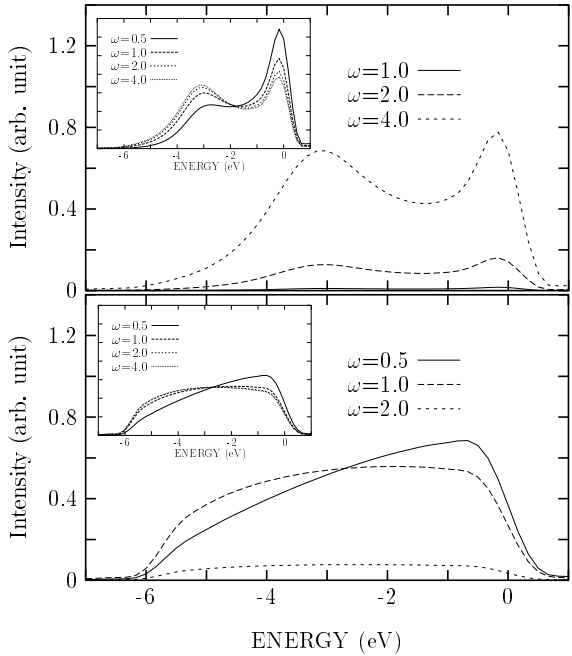}
\caption{The $\omega$-dependent valence PES are given for
$4f$-emission (upper panel) and $5d$-emission (lower panel).
The spectra at $\omega=0.5$ in the upper panel or $\omega=4.0$ in the lower 
will be so tiny that they are not illustrated. In the respective inset,
the spectra is normalized to have the same area to give the change of shape.
The unit of $\omega$ is the atomic unit (1 au = 27.2 eV).
}
\label{fig:w_separ}
\end{figure}

In Fig.\ref{fig:dipol_E}, we give the dipole matrix behaviors of
$\Delta_f(E)$ and $\Delta_d(E)$. This is obtained from very crude calculations,
but qualitatively quite consistent with Yeh and Lindau's\cite{Yeh}
calculation of photoionization cross sections for Ce.
We can simply expect from the behaviors of dipole elements the
general trend is that in the low energy, $d$-emission will
be dominant over $f$-emission, while in the high energy, $f$-emission
dominant over $d$-emission. We first show the calculation result
for separate $4f$ and $5d$ contributions with respect to
various photon energies $\omega$.

Figure \ref{fig:w_separ} shows $\omega$-dependent $4f$- and $5d$-emission,
the changes of spectral height and shape with $\omega$ varied.
The spectral height will be proportional to the square of 
dipole matrix elements at the corresponding energies and the shape
related to the behaviors of dipole element. If the $5d$ radial wave function 
does not vary significantly for La and Ce, the bottom panel of the figure
can be compared with the PES for La\cite{Wiel} and found to be
consistent with the experiment. We can also find as $\omega$ 
increases the spectral shape approaches the sudden approximation 
results (see the insets).

Now we investigate the total valence PES to which both $4f$- and $5d$-emission
contribute with respect to various photon energies $\omega$. 
To parameterize the relative strength of two dipole matrix effects, 
we define $\tilde{\Delta}_f$ and $\tilde{\Delta}_d$ as
$\tilde{\Delta}_f\equiv{\Delta_f(E)}|_{E=4.0}$,
$\tilde{\Delta}_d\equiv{\rm max}\{\Delta_d(\epsilon,E)\}$.
That is, $\tilde{\Delta}_f$ is defined as the value of $\Delta_f(E)$
at $E=4.0$ and $\tilde{\Delta}_d$ as the value of $\Delta_d(\epsilon,E)$
at $\epsilon=0.0$ and $E\approx 0.6$ (see Fig.\ref{fig:dipol_E}).
We give the behaviors of valence PES as $\omega$ varied 
for $|\tilde{\Delta}_d/\tilde{\Delta}_f|=2.0$
and $|\tilde{\Delta}_d/\tilde{\Delta}_f|=6.0$, respectively.

\begin{figure}
\vspace*{6.0cm}
\includegraphics{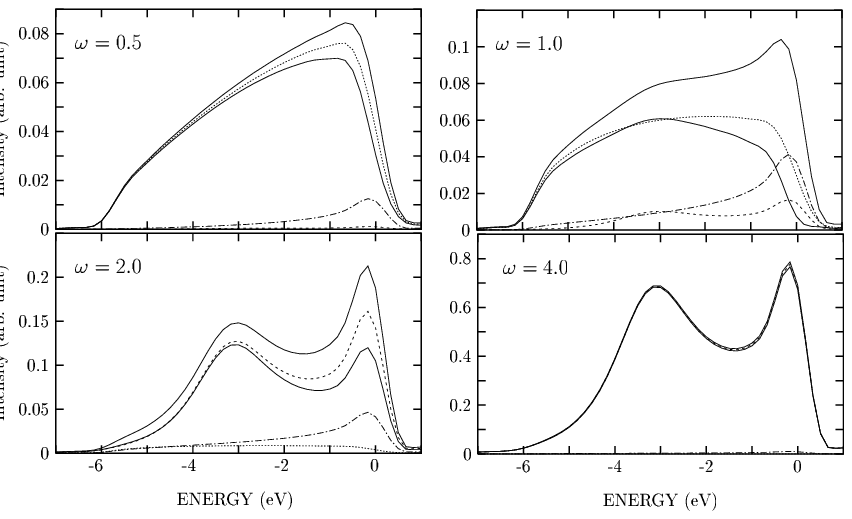}
\caption{The valence PES with respect to photon energies are provided.
The relative strength of two channels is taken as
$|\tilde{\Delta}_d/\tilde{\Delta}_f|=2.0$. Here the solid line is
the total spectra corresponding to constructive or destructive interference.
The dashed line represents $4f$-emission, the dotted line $5d$-emission,
and the dot-dashed line the interference contribution.
}
\label{fig:w_total2}
\end{figure}

\begin{figure}
\vspace*{6.0cm}
\includegraphics{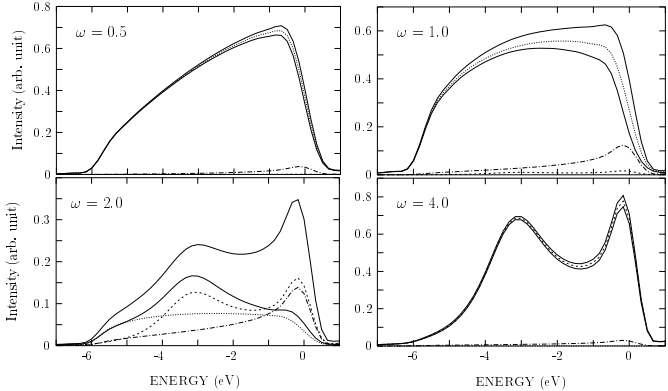}
\caption{The valence PES with respect to photon energies.
The relative strength of two channels is taken as 
$|\tilde{\Delta}_d/\tilde{\Delta}_f|=6.0$. Notations are same 
as in Fig.\ref{fig:w_total2}.
}
\label{fig:w_total6}
\end{figure}

In Figs.\ref{fig:w_total2} and \ref{fig:w_total6},
we see that at $\omega=0.5$ au,
the dominant contributions are from $5d$-band emission because
$\Delta_f(E)$ increases slowly compared to $\Delta_d(\epsilon,E)$,
however at $\omega=4.0$ au, most of the spectra in Ce arises from
the $4f$ electrons because $\Delta_d(\epsilon,E)$ rapidly falls off
over $E\sim 0.6$ au. Gunnarsson and Sch\"{o}nhammer\cite{Gunnar85}
have obtained the total emission spectra involving the interference
(of $4f$ and $5d$) based on the sudden approximation, 
but could not have discussed these behaviors with respect to $\omega$.
In the experiments, on the other hand, the increasing $4f$ 
and decreasing band features with varying $\omega$ has been used to separate 
the $4f$ structures\cite{Patthey}. That is, in the experiments, 
using He resonance lines, 
to subtract $\omega=0.78$ ($=21.2$ eV) result
from $\omega=1.5$ ($=40.8$ eV) result leads to approximately a $4f$-emission
for a moderate value of $|\tilde{\Delta}_d/\tilde{\Delta}_f|$ 
(i.e. say for $|\tilde{\Delta}_d/\tilde{\Delta}_f|\sim 2$).

Below, in Fig.\ref{fig:w_intf},
we can see the behaviors of interference as $\omega$ varies.
As $\omega$ increases, the interference become stronger at first and
then weaker again. And about $\omega=1.0$ or $2.0$ au, we see
the strong interference, where the spectra cannot be understood from
two separate emission spectra. Particularly in the case of
strong destructive interference, although the $4f$-emission always comprises
two peaks (see the inset of upper panel in Fig.\ref{fig:w_separ}),
the peak near the Fermi level may be smeared until $\omega$ becomes
a bit larger.

\begin{figure}
\vspace*{5.0cm}
\includegraphics{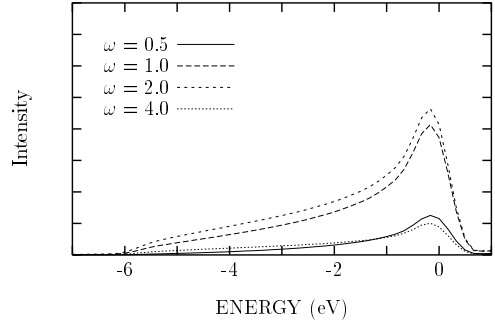}
\caption{The contribution of interferences are given
with respect to the photon energies. In the intermediate energies,
the interferences are very strong.
}
\label{fig:w_intf}
\end{figure}

\section{Low energy valence photoemission: effects of photoelectron scattering}

\begin{figure}
\vspace*{9.0cm}
\includegraphics{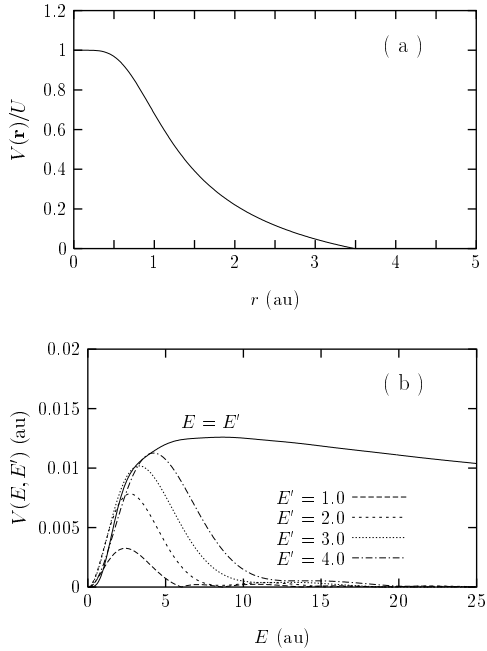}
\caption{(a) The photoelectron scattering potential $V({\bf{r}})$ is given
as normalized by $U$. (b) The diagonal and off-diagonal parts of 
scattering potential matrix are given.
}
\label{fig:Vmatrix}
\end{figure}

In the low photon energy region, we should consider the effects of 
photoelectron scattering potential induced by the hole left
by the electron emission as well as the dipole matrix behavior.
The band emission can also raise the shake-up effects like plasmon satellites
due to the fluctuation potential\cite{Arya}.
Nevertheless, in the present model, within $1/N_f$ expansion,
the relevant bases of Eqs.(\ref{II5})-(\ref{II10}) in the limit of 
$N_f\rightarrow\infty$ do not allow any conduction electron-hole 
excitation. Any shake-up behaviors from dielectric responses 
by band emissions then cannot be seen in the taken limit,
but in the next higher order of $1/N_f$.
For a hole in a localized $f$-level, however,
a small number of electrons may undergo measurable shifts
in response to the potential induced by a hole\cite{Lynch}.
For the photoelectron scattering potential, we should
go back to Eq.(\ref{II1}) and see the interaction of a
$f$-level impurity electron. In this section, we'll confine our discussion 
only to the $f$-level valence photoemission, i.e. here 
we do not consider the interference with band contributions.
In ${\cal H}$ ($={\cal H}_0+\Delta$), it needs noting that
the $f$-electron correlation is actually a quantity renormalized
by the conduction electrons, that is,
\begin{eqnarray}\label{V1}
U_{ff}n_fn_f&+&U_{fd}n_f\sum_{\nu}\int d\epsilon
\psi_{\epsilon\nu}^{\dagger}\psi_{\epsilon\nu}
\nonumber \\
&=&(U_{ff}-U_{fd})n_fn_f=Un_fn_f,
\end{eqnarray}
where we have used 
$n_f+\sum_{\nu}\int d\epsilon\psi_{\epsilon\nu}^{\dagger}\psi_{\epsilon\nu}$
is a conserved quantity. In the similar way, we can compose 
the scattering potential term $V_{SC}$ which must be added to ${\cal H}_0$
\bb\label{V2}
V_{\rm SC}=V_{4f}({\bf{r}})n_f+V_{5d}({\bf{r}})\sum_{\nu}\int d\epsilon
           \psi_{\epsilon\nu}^{\dagger}\psi_{\epsilon\nu}-V_{4f}({\bf{r}}),
\ee
where it should be noted that the initial neutral (ground) state is $4f^1$.
Then we have 
\begin{eqnarray}\label{V3}
V_{\rm SC}&=&[V_{4f}({\bf{r}})-V_{5d}({\bf{r}})]n_f-V_{4f}({\bf{r}})
\nonumber \\
          &=&[V_{4f}({\bf{r}})-V_{5d}({\bf{r}})](n_f-1)-V_{5d}({\bf{r}}).
\end{eqnarray}
We know $V_{5d}({\bf{r}})$ is much broader and weaker than $V_{4f}({\bf{r}})$
and better to be neglected. So we take $V_{\rm SC}$ as
\bb\label{V4}
V_{\rm SC}=V({\bf{r}})(n_f-1),
\ \ V({\bf{r}})=V_{4f}({\bf{r}})-V_{5d}({\bf{r}}).
\ee
Then we cut off the potential by taking the muffin-tin radius $r_{mt}$ as
the radius of neutral Ce atom, $r_{mt}=3.49$ au. 
$V_{4f}({\bf{r}})$ and $V_{5d}({\bf{r}})$ are evaluated from 
the Slater orbital. Now, $V({\bf{r}})$
is a short range one due to a screening of conduction $d$-electrons
and have a simple relation between $V({\bf{r}})$ and the intra-atomic
Coulomb correlation $U$,
\bb \label{V5}
U=\int d{\bf{r}}\rho_f({\bf{r}})V({\bf{r}})\approx V(0),
\ee
if we assume the $f$-level charge density $\rho_f({\bf{r}})$ and 
$f$-level is quite localized like the core level. Thus 
$V({\bf{r}})$ should be redefined by $\frac{1}{\varepsilon}V({\bf{r}})$,
where $\varepsilon$ is a dielectric constant chosen to make sure of 
Eq.(\ref{V5}), being due to screening by the surrounding,
so $\varepsilon$ is $V(0)/U\approx 5.24$.
The behavior of $V({\bf{r}})$ is given in Fig.\ref{fig:Vmatrix}(a).

Then we express the scattering potential in terms of 
the photoelectron basis function,
\bb\label{V6} 
V_{\rm SC}=\sum_{\nu}\int dEdE^{\prime}
       V(E,E^{\prime})\psi_{E\nu}^{\dagger}\psi_{E^{\prime}\nu}
       \left[\sum_{\nu^{\prime}}\psi_{\nu^{\prime}}^{\dagger}
                                \psi_{\nu^{\prime}}-1\right],
\ee
where the potential matrix element $V(E,E^{\prime})$ are
\bb\label{V7}
V(E,E^{\prime})=\int d{\bf{r}}\varphi_{E\nu}^{\ast}({\bf{r}})V({\bf{r}})
                \varphi_{E^{\prime}\nu}({\bf{r}}).
\ee
As in calculating the dipole matrix, 
the photoelectron basis function $\varphi_{E\nu}({\bf{r}})$
must be obtained by solving the Schr\"{o}dinger equation
under the atomic potential. But here we use simple
spherical Bessel function of $l=4$ as in the last section. We hopefully
expect the essential feature will not be spoiled by neglecting 
the phase shift $\delta(k)$.
So the desirable $\varphi_{E\nu}({\bf{r}})$ is
\bb\label{V8}
\varphi_{E\nu}({\bf{r}})=\frac{\sqrt{2}}{\sqrt{\pi}}(2E)^{1/4}
                         j_4(\sqrt{2E}r)Y_{4m}(\hat{\bf{r}}),
\ee
and the matrix element $V(E,E^{\prime})$ is
\bb\label{V9}
V(E,E^{\prime})=\frac{2}{\pi}(4EE^{\prime})^{1/4}\int dr r^2
                j_4(\sqrt{2E}r)V(r)j_4(\sqrt{2E^{\prime}}r),
\ee
whose explicit behaviors are shown in Fig.\ref{fig:Vmatrix}(b).

\begin{figure}
\vspace*{6.0cm}
\includegraphics{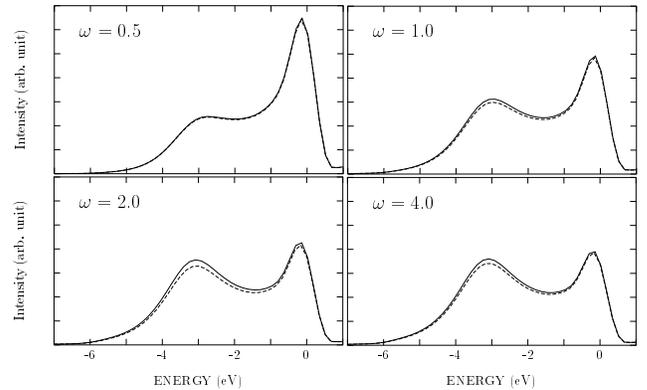}
\caption{Effects of scattering potential (solid line) are illustrated by
comparing with noninteracting results (dashed line) at given photon energies.
}
\label{fig:Veffect1}
\end{figure}

\begin{figure}
\vspace*{6.0cm}
\includegraphics{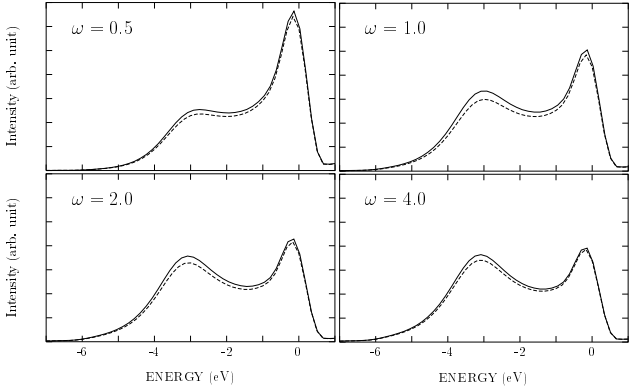}
\caption{Notations are same as in Fig.\ref{fig:Veffect1}. 
Here the scattering potential range has been made a bit larger
than in a Ce case, $r_{mt}=5.24$ au.
}
\label{fig:Veffect2}
\end{figure}

As $V_{SC}$ added, the total Hamiltonian ${\cal H}$ becomes
${\cal H}_0+\Delta+V_{SC}$. Under ${\cal H}$, the valence PES 
via $f$-channel can be calculated. The comparison of the results with $V_{SC}$ 
to those without $V_{SC}$ (still including the dipole elements) 
is provided in Fig.\ref{fig:Veffect1}.
The effects of scattering potential are quite small as shown in
Fig.\ref{fig:Veffect1}, which must be due to a weak scattering potential
in a typical Ce compounds.
Nevertheless, it's very meaningful to pursue a general consensus
in the valence PES about the photoelectron scattering effects.
In order to be more instructive, we also investigate the 
resulting behaviors for a slightly different potential whose
range is a bit larger by $50\%$ (see Fig.\ref{fig:Veffect2}).

In both Figs.\ref{fig:Veffect1} and \ref{fig:Veffect2}, it's notable
that there are no appreciable changes in the peak
near the Fermi level, while an increase in the ionization peak.
This will be understood from the scattering potential,
$V({\bf r})(n_f-1)$. The Fermi-level peak is from $4f^1\rightarrow 4f^1$,
which will not be affected much by the potential because of $n_f=1$,
but the ionization peak is from $4f^1\rightarrow 4f^0$.
It is important to grasp the underlying physics from the spectral 
changes as the potential range is increased from $r_{mt}=3.49$
to $r_{mt}=5.24$ ($1.5\times 3.49$).
Naturally a longer range potential results in more prominent effects
in the spectra. In two respects, 
the spectral changes due to the photoelectron scattering
look very similar to the core level PES in "shake down" case 
in the previous work of ours\cite{Lee}. 
First, if we simulate the absorption intensity ratio by 
the ratio of the ionization peak to the Fermi-level peak,
we find the constructive interference between intrinsic and 
extrinsic processes in the low energy regions just as in 
the shake down scenario of core level PES (the ratio is increased
due to the scattering). The second point is the relevant
energy scale governing the constructive interference due to
the photoelectron scattering. In Fig.\ref{fig:Veffect1}, 
we see the maximal scattering effects around $\omega\sim 2.0$,
while, in Fig.\ref{fig:Veffect2}, the energy scale giving the  
maximal effects is around $\omega\sim 1.0$. That is, 
the governing energy scale is decreased by an increased range.
In the core level PES, we found the relevant energy ($\tilde{E}$) can be 
directly related by the potential range ($\tilde{R}$) as 
$\tilde{E}=1/(2\tilde{R}^2)$, which tempts the application of 
the criteria to the present system. 
And it then can be a natural motivation that
we make an analysis for the adiabatic-sudden transition
in this system and try to answer if the criteria found previously in core
level PES can be still valid in this valence PES or not.
This question is extensively discussed in the next section, where
we propose the simplified two-level model for the sake of simplicity.

\section{adiabatic-sudden transition for 2-electron and 
$N_{\lowercase{f}}=2$}

The AIM can be reduced into the two-level model, i.e.
the whole continuum band is replaced by one level.
The Hamiltonian ${\cal H}_0$ we should now consider is
\begin{eqnarray}\label{VI1}
{\cal H}_0&=&\epsilon_d\sum_{\sigma}\psi_{d\sigma}^{\dagger}\psi_{d\sigma}
          +\epsilon_f\sum_{\sigma}\psi_{f\sigma}^{\dagger}\psi_{f\sigma}
\nonumber \\
          &+&V\sum_{\sigma}(\psi_{f\sigma}^{\dagger}\psi_{d\sigma}
                         +\psi_{d\sigma}^{\dagger}\psi_{f\sigma})
          +Un_{f\uparrow}n_{f\downarrow},
\end{eqnarray}
where $n_{f\sigma}=\psi_{f\sigma}^{\dagger}\psi_{f\sigma}$. Then 
the analogy of the present problem with the core level PES 
for the shake down case (having "level crossing" as the hole
is created) is more evident.
The change of relevant electronic levels is given schematically before 
and after the photoemission in the following Fig.\ref{fig:schematic}.

\begin{figure}
\vspace*{3.3cm}
\includegraphics{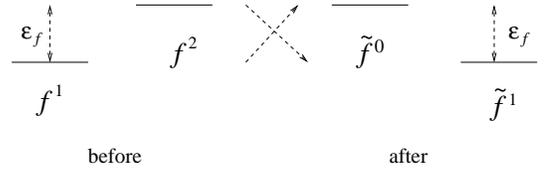}
\caption{Schematic view of the relevant configurations
in initial and final stage.
Here we assume $2\epsilon_f+U\approx 0$. Note there's a level crossing
before and after the emission.
}
\label{fig:schematic}
\end{figure}

We can introduce three states $|f^0\rangle$, $|f^1\rangle$, and $|f^2\rangle$
as follows,
$$
|f^0\rangle=\psi_{d\uparrow}^{\dagger}\psi_{d\downarrow}^{\dagger}|0\rangle,
$$
$$
|f^1\rangle=\frac{1}{\sqrt{2}}
        [\psi_{d\uparrow}^{\dagger}\psi_{f\downarrow}^{\dagger}|0\rangle
        -\psi_{d\downarrow}^{\dagger}\psi_{f\uparrow}^{\dagger}|0\rangle],
$$
$$
|f^2\rangle=\psi_{f\uparrow}^{\dagger}\psi_{f\downarrow}^{\dagger}|0\rangle,
$$
then we can express ${\cal H}_0$ in these bases,
\bb\label{VI2}
{\cal H}_0=\left (\begin{array}{ccc}
           0 & \bar{V} & 0 \\
           \bar{V} & \Delta\epsilon & \bar{V} \\
           0 & \bar{V} & 2\Delta\epsilon+U
           \end{array}  \right) + 2\epsilon_d,
\ee
where $\Delta\epsilon=\epsilon_f-\epsilon_d$ and $\bar{V}=\sqrt{2}V$.
For simplicity, we will put $2\Delta\epsilon+U=0$.
The ground state corresponding to the initial state of photoemission,
$|\Psi_0\rangle$ is
\bb\label{VI3}
|\Psi_0\rangle=\frac{\bar{V}}{\sqrt{\Delta_0^2+2\bar{V}^2}}
               \left[|f^0\rangle+\frac{\Delta_0}{\bar{V}}|f^1\rangle
               +|f^2\rangle\right],
\ee
where $\Delta_0=\frac{1}{2}\Delta\epsilon
-\frac{1}{2}\sqrt{\Delta\epsilon^2+8\bar{V}^2}$ and its energy $E_0$ is
\bb\label{VI4}
E_0=\Delta_0+2\epsilon_d.
\ee
The final states of the target are given by the following another set of bases,
$$
|\tilde{f}^0;\sigma\rangle=\psi_{d\sigma}^{\dagger}|0\rangle,
$$
$$
|\tilde{f}^1;\sigma\rangle=\psi_{f\sigma}^{\dagger}|0\rangle,
$$
then the Hamiltonian $\tilde{\cal H}_0$ with one $f$-electron emitted is
\bb\label{VI5}
\tilde{\cal H}_0=\left(\begin{array}{cc}
           0 & \bar{V}/\sqrt{2}\\
           \bar{V}/\sqrt{2} & \Delta\epsilon
           \end{array} \right) +\epsilon_d.
\ee
Note $\psi_{f\sigma}|f^0\rangle=0$,
$\psi_{f\sigma}|f^1\rangle=\frac{1}{\sqrt{2}}\sigma|\tilde{f}^0;-\sigma\rangle$,
and $\psi_{f\sigma}|f^2\rangle=\sigma|\tilde{f}^1;-\sigma\rangle$
and $\sigma$ is $\pm 1$.
Here $\sigma$ is actually a redundant parameter.
The possible final target states will be given by the eigenstates of
$\tilde{\cal H}_0$ in Eq.(\ref{VI5}),
\bb\label{VI6}
|\Psi_1;\sigma\rangle=\cos\varphi|\tilde{f}^0;\sigma\rangle
                     -\sin\varphi|\tilde{f}^1;\sigma\rangle,
\ee
\bb\label{VI7}
|\Psi_2;\sigma\rangle=\sin\varphi|\tilde{f}^0;\sigma\rangle
                     +\cos\varphi|\tilde{f}^1;\sigma\rangle,
\ee
where $E_{1\atop 2}=\frac{1}{2}\Delta\epsilon
                 \mp\frac{1}{2}\sqrt{\Delta\epsilon^2+2\bar{V}^2}+\epsilon_d$
($\delta E=\sqrt{\Delta\epsilon^2+2\bar{V}^2}$),
and the parameter $\varphi$ ($\frac{\pi}{4}<\varphi<\frac{\pi}{2}$)
is determined by
\bb\label{VI8}
\tan\varphi=\frac{1}{\sqrt{2}}(\sqrt{w^2+2}-w),
\mbox{ }\mbox{ }w=\frac{\Delta\epsilon}{\bar{V}}.
\ee
Also noticeable is
\bb\label{VI9}
\psi_{f\sigma}|\Psi_0\rangle=\frac{1}{\sqrt{2}}
     \left[-\sin\theta|\tilde{f}^0;-\sigma\rangle
           +\cos\theta|\tilde{f}^1;-\sigma\rangle\right],
\ee
where the parametric angle $\theta$ ($\frac{\pi}{4}<\theta<\frac{\pi}{2}$) is
\bb\label{VI10}
\cot\theta=\frac{1}{2\sqrt{2}}(\sqrt{w^2+8}+w).
\ee
Now we consider the optical activation Hamiltonian $\Delta$,
\bb\label{VI11}
\Delta=\sum_{k\sigma}M_k\psi_{k\sigma}^{\dagger}\psi_{f\sigma},
\mbox{ }\mbox{ }
\left(\mbox{or}\ \Delta=\int dE\Delta(E)\psi_{E\sigma}^{\dagger}
\psi_{f\sigma}\right).
\ee
Within the first order perturbation theory,
the photoemission matrix element $M(i,k\sigma)$ ($i=1,2$) will be
\begin{eqnarray}\label{VI12}
& &M(i,k\sigma)\nonumber \\
&=&\langle\Psi_i;-\sigma|\psi_{k\sigma}\left[
             1+V\frac{1}{E-{\cal H}_0-T+i\eta}\right]
             \Delta|\Psi_0\rangle \nonumber \\
&=&m_iM_k+\sum_jc_{ij}m_j
  \sum_{k^{\prime}}\frac{V_{kk^{\prime}}M_{k^{\prime}}}
  {\omega+E_0-E_j-\frac{1}{2}{k^{\prime}}^2+i\eta},
\end{eqnarray}
where the scattering potential $V_{SC}$ is taken as
\bb\label{VI13}
V_{SC}=V({\bf r})(n_f-1),\ \
V_{kk^{\prime}}=\int d{\bf r}\varphi_{k\sigma}^{\ast}({\bf r})V({\bf r})
                            \varphi_{k^{\prime}\sigma}({\bf r})
\ee
and thus
$$
m_i=\langle\Psi_i;-\sigma|\psi_{f\sigma}|\Psi_0\rangle,
\mbox{ }\mbox{ }
c_{ij}=\langle\Psi_i;-\sigma|n_f|\Psi_j;-\sigma\rangle-\delta_{ij}.
$$
That is, the coefficients are
$$
m_1=-\frac{1}{\sqrt{2}}\sin(\varphi+\theta)\sigma,
\mbox{ }\mbox{ }
m_2=\frac{1}{\sqrt{2}}\cos(\varphi+\theta)\sigma,
$$
$$
c_{11}=-\cos^2\varphi, \mbox{ }\mbox{ }
c_{22}=-\sin^2\varphi, \mbox{ }\mbox{ }
c_{12}=c_{21}=-\sin\varphi\cos\varphi.
$$
If we consider a ratio between the main and the satellite 
absorption intensity divided by a noninteracting case
$r(\omega)/r_0(\omega)$, $r(\omega)/r_0(\omega)$ is
\begin{eqnarray}\label{VI14}
& &\frac{r(\omega)}{r_0(\omega)} \\ \nonumber
&=&\left |
   \frac{1+\sin^2\varphi\frac{\tilde{V}}{\tilde{E}}F_{k_2}
           \left(\frac{\tilde{\omega}}{\tilde{E}}\right)
          -\frac{\sin 2\varphi\sin(\varphi+\theta)}{2\cos(\varphi+\theta)}
           \frac{\tilde{V}}{\tilde{E}}F_{k_2}
           \left(\frac{\tilde{\omega}+\delta E}{\tilde{E}}\right)}
          {1+\cos^2\varphi\frac{\tilde{V}}{\tilde{E}}F_{k_1}
           \left(\frac{\tilde{\omega}+\delta E}{\tilde{E}}\right)
          -\frac{\sin2\varphi\cos(\varphi+\theta)}{2\sin(\varphi+\theta)}
           \frac{\tilde{V}}{\tilde{E}}F_{k_1}
           \left(\frac{\tilde{\omega}}{\tilde{E}}\right)}
      \right|^2,
\end{eqnarray}
where $k_{1\atop 2}=\sqrt{2(\omega+E_0-E_{1\atop 2})}$,
$\tilde{\omega}=E_2-E_0$ (threshold energy for the satellite), and 
we have extended the model matrix elements 
$M_k$ and $V_{kk^{\prime}}$ used in the previous core level case,
\bb\label{VI15}
\sum_{k^{\prime}}\frac{V_{kk^{\prime}}M_{k^{\prime}}}
                      {\epsilon-\epsilon_{k^{\prime}}+i\eta}
    =-\frac{\tilde{V}}{\tilde{E}}M_kF_k(\epsilon/\tilde{E}),
\ee
\bb\label{VI16}
F_k(\epsilon)=\frac{1}{\pi}\int_0^{\infty}
              \frac{x^{10}dx}{[1+x^5]^2\left[1+(\tilde{R}k-x)^2\right]
              [x^2-\epsilon-i\eta]},
\ee
\bb\label{VI17}
M_k=\frac{(\tilde{R}k)^5}{1+(\tilde{R}k)^5},
\ee
\bb\label{VI18}
V_{kk^{\prime}}=\frac{\tilde{V}\tilde{R}}{R}
                \frac{(\tilde{R}^2kk^{\prime})^5}
                     {[1+(\tilde{R}k)^5][1+(\tilde{R}k^{\prime})^5]
                      [1+\tilde{R}^2(k-k^{\prime})^2]},
\ee
where $\tilde{R}$ is the characteristic length scale of the system
directly related to the potential range and $\tilde{E}=1/2\tilde{R}^2$.
Here it is found that from Eq.(\ref{VI14}), $r(\omega)/r_0(\omega)$
can be written essentially in the same mathematics as in the core level case.
In the following Fig.\ref{fig:F}, we give the behaviors of $F(\epsilon)$.

\begin{figure}
\vspace*{5.0cm}
\includegraphics{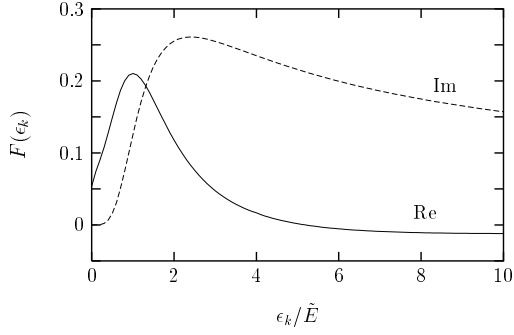}
\caption{The function $F(\epsilon_k)$ defined in Eq.(\ref{VI16}). 
Both the real and imaginary parts are given.
}
\label{fig:F}
\end{figure}

\begin{figure}
\vspace*{5.0cm}
\includegraphics{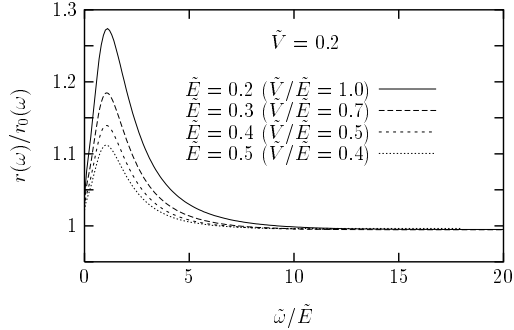}
\caption{The ratio $r(\omega)/r_0(\omega)$ as a function of 
$\tilde{\omega}/\tilde{E}$ for several values of $\tilde{V}/\tilde{E}$.
$\Delta\epsilon/\bar{V}=-2.0$ is taken.
}
\label{fig:ptb}
\end{figure}

Similarly to the core case, we always have an overshoot 
behavior in $r(\omega)/r_0(\omega)$ in the low energy limit,
when $\delta E=0$,
\bb\label{VI19}
\frac{r(\omega_{th})}{r_0(\omega_{th})}
                   =\left[\frac{1-F(0)\frac{\tilde{V}}{\tilde{E}}
                          \frac{\sin\varphi\sin\theta}{\cos(\varphi+\theta)}}
                               {1+F(0)\frac{\tilde{V}}{\tilde{E}}
                          \frac{\cos\varphi\sin\theta}{\sin(\varphi+\theta)}}
                    \right]^2 > 1,
\ee
where $F(0)=0.052286$ and $\frac{\pi}{4}<\varphi<\frac{\pi}{2}$,
$\frac{\pi}{4}<\theta<\frac{\pi}{2}$, and
$\frac{\pi}{2}<\varphi+\theta<\pi$ should be noted.
In Fig.\ref{fig:ptb}, we show $r(\omega)/r_0(\omega)$ as a function of 
$\tilde{\omega}/\tilde{E}$ for a few values of $\tilde{V}/\tilde{E}$
($\tilde{\omega}\equiv\omega-\omega_{th}$).

Fig.\ref{fig:ptb} shows as $\omega$ increases the ratio also increases
and reaches a maximum. From the arguments
in our previous work\cite{Lee}, we see roughly $\tilde{R}\sim R_0/3$
and $\tilde{V}\sim 3V(0)/2$ and for Ce, $\tilde{R}\sim 1$
and $\tilde{V}\sim 0.3$, which leads to
$\tilde{V}/\tilde{E}\sim 0.6$. 
Most notable is that the curves have the universal 
feature independent of $\tilde{V}/\tilde{E}$, i.e. the overshoots  
disappear at about $\tilde{\omega}/\tilde{E}\sim 10$ 
in all relevant parameter region. 
This means the adiabatic-sudden transition depends only on $\tilde{E}$, 
that is, $\tilde{R}$, even if the amplitude of overshoot is 
from $\tilde{V}/\tilde{E}$.
Beyond the first order perturbation,
the overshoot range will be reduced due to the multiple scattering,
but the universal behavior not changed.
This conclusion is exactly identical to 
that in the core level case\cite{Lee} and implies the same
criteria can be applied also to the valence PES case.

\section{Conclusion}

We have studied the valence photoemission spectra in the Anderson
impurity model aiming at Ce compounds. Using the time-dependent
formulation and $1/N_f$ expansion, we can treat the problem 
exactly up to ${\cal O}(1/N_f)^0$. For Ce compounds, $N_f=\infty$
can be a good approximation for $N_f=14$. Within the formalism,
to evaluate the photoemission spectra is corresponding to 
solving the time-dependent Schr\"{o}dinger equation.

To investigate the low energy photoemission spectra, we should
consider the dipole matrix and photoelectron scattering matrix
additionally compared to the sudden approximation valid in high energy limit.
In view of experiment, the valence PES always consist of $f$-emission and
band emission. So we considered both dipole matrix elements
having explicit $E$-dependencies and obtained the total spectra
as well as two separate spectra with respect to the photon energies.
The relative strength and sign of two dipole elements are crucial
in the total spectra. Due to differences in the energy scales
of $4f$ and $5d$ dipole elements, the general trends are that,
in a very low energy ($\omega\lesssim 1.0$ au), the $5d$ emission
is dominant, while the $4f$ emission increases and dominates over 
the $5d$ emission in a high energy ($\omega\gtrsim 2.0$ au).
It is also found that the interference effects of
$f$-emission and band emission (from $d$) are highly nontrivial
especially in the intermediate energy region. 
In the intermediate region 
($\omega\sim 1.0$ or 2.0 au depending on the relative strength), 
due to a strong peak of the interferences near the Fermi level,
the Kondo resonance peak of $4f$ emission may be smeared 
in the case of destructive interference. The constructive or destructive 
interference will be determined by the relative sign.

We also studied the effects of scattering potential for the
$f$-electron emission. Scattering effects on the band emission 
is neglected in the infinite $N_f$ limit.
The potential matrix also includes the kinetic energy dependencies. 
It's a general result that, in the case where the photoelectron 
couples to the localized excitation, the arrival at the sudden transition
is much faster compared to the extended excitation.
Effects of scattering potential gives the similar spectral 
changes to the core level PES in shake down case if we assign 
the Kondo resonance and ionization peak to the main and satellite
peak, respectively. 
Behaviors of two-peak ratio is reminiscent of the previous analysis of 
core level PES. Therefore, this can be a motivation to do 
a further analysis for the adiabatic-sudden behavior.
We can then cast a question, {\it "The criteria found in core level case 
can be also valid in the valence case?"} 
To explore it, first we simplify the model into just a two-level one,
where the whole conduction band is replaced by one level.
Then it is found the intensity ratio of two peaks
can be written actually in the same way
as in the core case. Through the same analysis, we can find that 
also in the valence PES, the sudden transition happens on the 
energy scale of $\tilde{E}=1/(2\tilde{R}^2)$, where $\tilde{R}$ is a 
typical length scale of the scattering potential.

Finally, we would like to make it clear the limit or shortcomings of
our present model. Within the model, we cannot yet arrive at 
the realistic photoemission in a true solid. The AIM includes 
the continuum band describing a crystal, which has a surface.
Then the position of a surface with respect to the created hole
will enter as an important parameter. In a true solid, we never reach
a vanishing extrinsic scattering because the corresponding cross section
goes to zero, while the photoelectron is emitted from an infinite depth.
And to correctly predict 
the amount of photocurrent, we also should use the damped wavefunctions,
not undamped partial wave. These additional extrinsic effects 
may be as important as the hole-induced scattering potential treated here.
Therefore, it should be understood that the scope of this paper
is to explore the effects of hole-induced scattering potential 
in the valence PES in AIM describing the Ce compounds
apart from additional ingredients.

\section*{acknowledgement}

The author would like to appreciate Lars Hedin, Olle Gunnarsson, ByungIl Min 
for valuable discussions and their critical reading the manuscript.

\end{multicols}

\end{document}